\documentclass[aps,prl,twocolumn,superscriptaddress,amsfonts,amsmath,amssymb,showpacs,floatfix,nofootinbib]{revtex4-1}

\usepackage{graphicx}
\usepackage{longtable}
\usepackage{bm}
\usepackage{multirow}
\usepackage{hyperref}
\usepackage{ulem}
\hypersetup{colorlinks,citecolor=blue,filecolor=blue,linkcolor=blue,urlcolor=blue}
\usepackage{color}
\usepackage{dcolumn}
\usepackage{tabularx}
\usepackage[usenames,dvipsnames]{xcolor}
\usepackage{subfigure}

\definecolor{awesome}{rgb}{1.0, 0.13, 0.32}
\newcommand{\iso}[2]{$^{#1}$#2}
\newcommand{\orb}[3]{$#1#2_{#3/2 }$}

\newcommand{\dHe}{($d$,$^3$He)}

\begin{document}

\title{Study of the Isomeric State in \iso{16}{N} Using the \iso{16}{N}$^{g,m}$(\bf{\textit{d}},$^3$He)~Reaction}

\author{T.~L.~Tang}
\email{rtang@fsu.edu}
\altaffiliation[Present address: ]{Department of Physics, Florida State University, Tallahassee, Florida 32306, USA}
\affiliation{Physics Division, Argonne National Laboratory, Argonne, Illinois 60439, USA}
\author{C.~R.~Hoffman}
\author{B.~P.~Kay}
\affiliation{Physics Division, Argonne National Laboratory, Argonne, Illinois 60439, USA}
\author{I.~A.~Tolstukhin}   \affiliation{Physics Division, Argonne National Laboratory, Argonne, Illinois 60439, USA}
\author{S.~Almaraz-Calderon} \affiliation{Physics Department, Florida State University, 600 W College Ave, Tallahassee, FL 32306, USA}
\author{B.~W.~Asher}     \affiliation{Physics Department, Florida State University, 600 W College Ave, Tallahassee, FL 32306, USA}
\author{M.~L.~Avila} \affiliation{Physics Division, Argonne National Laboratory, Argonne, Illinois 60439, USA}
\author{Y.~Ayyad}   \affiliation{National Superconducting Cyclotron Laboratory, Michigan State University, 640 S Shaw Ln, East Lansing, MI 48824, USA}
\author{K.~W.~Brown}   \affiliation{National Superconducting Cyclotron Laboratory, Michigan State University, 640 S Shaw Ln, East Lansing, MI 48824, USA}
\author{D.~Bazin}   \affiliation{National Superconducting Cyclotron Laboratory, Michigan State University, 640 S Shaw Ln, East Lansing, MI 48824, USA}
\author{J.~Chen} \altaffiliation[Present address: ]{Physics Division, Argonne National Laboratory, Argonne, Illinois 60439, USA}  \affiliation{National Superconducting Cyclotron Laboratory, Michigan State University, 640 S Shaw Ln, East Lansing, MI 48824, USA}
\author{K.~A.~Chipps}   \affiliation{Oak Ridge National Laboratory, 1 Bethel Valley Rd, Oak Ridge, TN 37830, USA}
\author{P.~A.~Copp}        \affiliation{Physics Division, Argonne National Laboratory, Argonne, Illinois 60439, USA}
\author{M.~Hall}   \affiliation{Oak Ridge National Laboratory, 1 Bethel Valley Rd, Oak Ridge, TN 37830, USA}
\author{H.~Jayatissa} \affiliation{Physics Division, Argonne National Laboratory, Argonne, Illinois 60439, USA}
\author{H.~J.~Ong}  \affiliation{Research Center of Nuclear Physics, 10-1 Mihogaoka, Ibaraki, Osaka, 567-0047 , Japan}
\author{D.~Santiago-Gonzalez} \affiliation{Physics Division, Argonne National Laboratory, Argonne, Illinois 60439, USA}
\author{D.~K.~Sharp}  \affiliation{Department of Physics, University of Manchester, M13 9PL Manchester, United Kingdom}
\author{J.~Song}   \affiliation{Physics Division, Argonne National Laboratory, Argonne, Illinois 60439, USA}
\author{S.~Stolze}      \affiliation{Physics Division, Argonne National Laboratory, Argonne, Illinois 60439, USA}
\author{G.~L.~Wilson} \affiliation{Department of Physics and Astronomy, Louisiana State University, LA 70803, USA} \affiliation{Physics Division, Argonne National Laboratory, Argonne, Illinois 60439, USA}
\author{J.~Wu}  \altaffiliation[Present address: ]{National Superconducting Cyclotron Laboratory, Michigan State University, 640 S Shaw Ln, East Lansing, MI 48824, USA} \affiliation{Physics Division, Argonne National Laboratory, Argonne, Illinois 60439, USA}

\date{\today}

\begin{abstract}
The isomeric state of \iso{16}{N} was studied using the \iso{16}{N}$^{g,m}$\dHe~proton-removal reactions at \mbox{11.8~MeV/$u$} in inverse kinematics. The \iso{16}{N} beam, of which 24\% was in the isomeric state, was produced using the ATLAS in-fight system and delivered to the HELIOS spectrometer, which was used to analyze the \iso{3}{He} ions from the ($d$,\iso{3}{He}) reactions. The simultaneous measurement of reactions on both the ground and isomeric states, reduced the systematic uncertainties from the experiment and in the analysis. A direct and reliable extraction of the relative spectroscopic factors was made based on a Distorted-Wave Born Approximation approach. The experimental results suggest that the isomeric state of \iso{16}{N} is an excited neutron-halo state.
The results can be understood through calculations using a Woods-Saxon potential model, which captures the effects of weak-binding.



\end{abstract}


\maketitle

\section{Introduction}

Many nuclei have been found to exhibit halo-like properties over the past three decades~\cite{Tanihata2013}. Although there is no precise definition for a nuclear halo, the general consensus is that a nucleus in a halo state has a large root mean square (rms) matter radius, as a consequence of the spatially extended nature of the valence nucleon(s)~\cite{Jensen2004, Tanihata1996, Nakamura2007, Nortershauser2009,Marques2002}. A single-neutron halo state can be linked to the weak-binding effect~\cite{Hoffman2014, Hoffman2014_2, Hoffman2016}, that any weakly bound state that is dominated by an $s$-orbital configuration and in proximity to the neutron threshold will have a large rms matter radius. 

The weak binding effect states the following: the $s$-orbital is special in that the centripetal force is absent, hence, a neutron experiences no barrier as its energy approaches the binding-energy threshold, S$_n\approx0$. When the $s$-orbital neutron is weakly bound, less than $\sim$2 ~MeV away from the threshold, the effective potential radius is large or similarly the potential becomes shallow. This allows the orbital to spatially spread outside the potential, i.e. the orbital has a significantly large rms radius. 


One of the most well known $s$-shell neutron halo states is the ground state of \iso{15}{C} \cite{Fang2004}. The experimental \iso{15}{C}--\iso{12}{C} scattering data shows a larger reaction cross section when compared to its neighboring carbon isotopes. The extracted longitudinal momentum distribution is also narrow as expected from an s-orbital~\cite{Fang2004}. 
The neutron spectroscopic factors (SFs) of the \orb{1}{s}{1} and \orb{0}{d}{5} orbitals were found to be 0.88 and 0.62, respectively, using the \iso{14}{C}($d$,$p$) reaction~\cite{Goss1975}. The SF of the \orb{1}{s}{1} orbital was found to be 0.9 using a knockout reaction on \iso{15}{C}~\cite{Terry2004}. Similarly, a neutron adding reaction on \iso{15}{N} has shown that the SFs of the \orb{1}{s}{1} are slightly larger than that of the \orb{0}{d}{5} orbital~\cite{Bardayan2009}. In both cases, the SFs for the $\ell=0$ neutron seem to be larger and closer to 1. However, a direct comparison between these SFs is not trivial due to a model uncertainty of around 20\% coming from comparing different $\Delta \ell$ transfers in the Distorted-Wave Born Approximation (DWBA) calculation~\cite{Ben2013}. 

This picture of halo states applies to excited states as well~\cite{Otsuka1993}, for example, excited halo states have been found or suggested in \iso{6}{Li}~\cite{Li2002}, \iso{11}{Be}~\cite{Belyaeva2014}, \iso{12}{Be}~\cite{Ritu2010}, \iso{12}{B}~\cite{Liu2001}, and \iso{13}{C}~\cite{Belyaeva2014, Liu2001}. 
The isomeric state of \iso{16}{N} also has the underlying makeup to form an excited halo state. The single-particle configuration of the 0.12 MeV ($J^\pi=0^-$) isomeric state of \iso{16}{N} is dominated by the \mbox{${^{14}\textrm{C}} \otimes \pi 0p_{1/2} \otimes \nu 1s_{1/2}$} configuration~\cite{Bardayan2009,Fang2004, Yasue2016, Goss1975, Tilley1993, Guo2014, Mairle1978, Terry2004}. The effective single-particle energy (ESPE) of the $s$-orbital is \mbox{-2.16 MeV}~\cite{Bardayan2009, Guo2014, Wang2021}. 
Ref.~\cite{Li2016} used the asymptotic normalization coefficients (ANCs) to deduce the rms radius of the s-orbital in the $0^-$ level and the result was $4.82\pm0.42$~fm, which is 53\% larger than the conventional size of the nucleus $\sim1.25A^{1/3} = 3.15$~fm. 
Recently, larger one-neutron removal cross section were observed for a \iso{16}{N} beam consisting of $\sim 24\%$ \iso{16}{N}$^m$ as compared to a \iso{16}{N} beam with only $\sim9\%$ isomer content~\cite{Fukutome2022}.

\begin{figure}[ht]
\centering
\includegraphics[trim=3.1cm 2.8cm 0cm 2.6cm, clip, width=8.5cm]{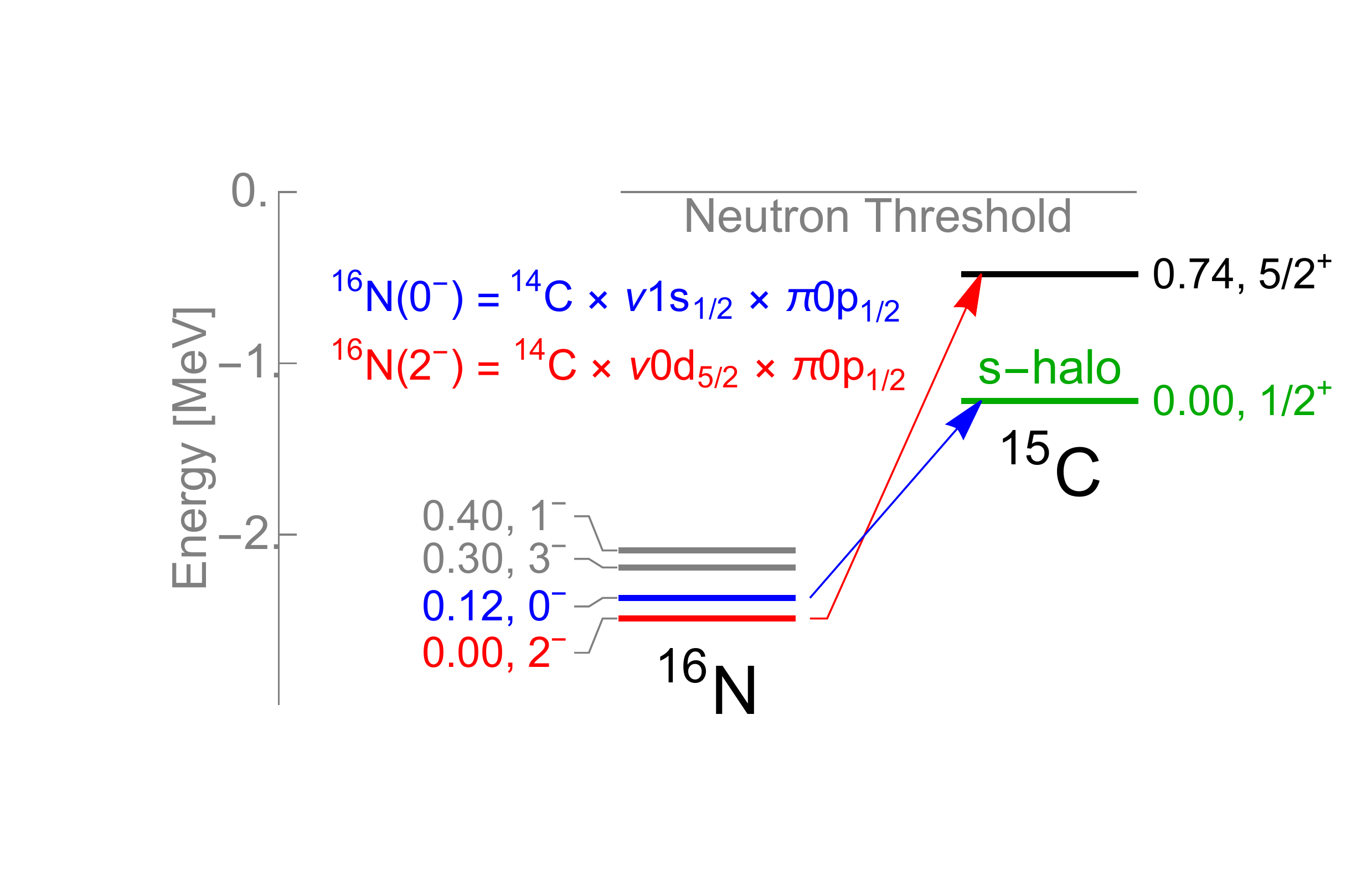}
\caption{\label{fig1} 
The energy levels of \iso{16}{N} and \iso{15}{C}. 
The two arrows indicate the transition for the \orb{0}{p}{1} proton removal.}
\end{figure}

In order to study the relative overlap between initial neutron \orb{1}{s}{1} and \orb{0}{d}{5} states in \iso{16}{N} and their counterparts in \iso{15}{C}, the simultaneous measurement of the \iso{16}{N}$^{g,m}$\dHe\iso{15}{C} reactions were performed in inverse kinematics~(Fig.\ref{fig1}). The simultaneous measurement reduced the experimental systematic uncertainty and suppressed the DWBA uncertainty as both reactions had the same $\ell=1$ proton removal at the same momentum matching, and regions of identical acceptance. The neutron structure between the initial and final states may be inferred by comparing the relative SFs between these two proton-removal reactions. Using proton removal to infer the neutron-shell structure was demonstrated in the \iso{25}{F}($p$,$2p$) reaction~\cite{tang2020}.

The \iso{14}{C} ($J^\pi=0^+$) ground state is a good core of \iso{15}{C} \cite{Yasue2016,Goss1975, Terry2004} and \iso{15}{N}~\cite{nucleiA13_15}. The configuration of the \iso{15}{C} ground state is dominated by $[$\iso{14}{C}$\otimes\nu$\orb{1}{s}{1}$]_{J^\pi = 1/2^+}$ and the first excited state is dominated by  $[$\iso{14}{C}$\otimes\nu$\orb{0}{d}{5}$]_{J^\pi=5/2^+}$. The configuration of the \iso{15}{N} ground state is dominated by $[$\iso{14}{C}$\otimes\pi$\orb{0}{p}{1}$]_{J^\pi = 1/2^-}$. With the \iso{15}{C} and \iso{15}{N} single-particle configurations, the configuration of the \iso{16}{N} ground state is dominated by $[$\iso{15}{C}($5/2^+$)$\otimes\pi$\orb{0}{p}{1}$]_{J^\pi=2^-}$, and the \iso{16}{N} isomeric state is dominated by $[$\iso{15}{C}($1/2^+$)$\otimes\pi$\orb{0}{p}{1}$]_{J^\pi=0^-}$. This configuration of \iso{16}{N} was verified in Refs.~\cite{Bardayan2009, Tilley1993, Guo2014,  Mairle1978}.

The SFs were extracted from the \iso{16}{N}\dHe\iso{15}{C} reaction via the ratio between the measured yields and those calculated by DWBA. The SF are a measure of the overlap between the initial and final wavefunctions. Therefore, the ratio of the SFs for proton removal from the ground and isomeric state in \iso{16}{N}, reflects how similar the two overlaps are with each other. This quantity, $R$, extracted in the present work, may be expressed as
\begin{equation}\label{eq:1}
\begin{aligned}
R= \frac{(SF)^g_{0p_{1/2}}}{(SF)^m_{0p_{1/2}}} &=\frac{\alpha^g(0p_{1/2})}{\alpha^m(0p_{1/2})} \frac{\beta^g(0d_{5/2})}{\beta^m(1s_{1/2})}\\
&\approx \left(1\right)\frac{1}{\beta^m(1s_{1/2})} \gtrsim 1. 
\end{aligned}
\end{equation}
Here, the spectroscopic overlap breaks down into proton and neutron parts.
The $\alpha^{g/m}(\pi)$ [$\beta^{g/m}(\nu)$] represents a measure of the similarity between the valence $\pi$ proton ($\nu$ neutron) wavefunctions of \iso{16}{N} and \iso{15}{C} for the \mbox{\iso{16}{N}$^{g/m}$($d$,\iso{3}{He})} reaction. In the next line of the equation, three assumptions are made: 1) \iso{14}{C} can be considered as a good core of \iso{15}{C} and \iso{16}{N}~\cite{Yasue2016, Bedoor2016, nucleiA13_15}; 2) the two $\alpha^{g/m}$ values from the proton removal reactions can be assumed to be equal as no change in the proton \orb{0}{p}{1} occupancy is expected between the different states in \iso{15}{C} and \iso{16}{N}. 
Thus, the neutron parts can be inferred from the measured ratio $R$; 3) The $\beta(0d_{5/2})$ should be close to unity as the states comprising of the neutron $d$-orbitals in \iso{16}{N} and \iso{15}{C} should also be similar. 
Therefore, $R$ should be $\gtrsim 1$, and is dependent on the overlap or similarities between the \orb{1}{s}{1} neutrons in the \iso{16}{N} isomer and \iso{15}{C} halo ground state. 


When the \orb{0}{p}{1} proton is removed from the \iso{16}{N} ground (isomeric) state of $J^\pi = 2^-$ ($0^-$), the $5/2^+$ excited ($1/2^+$ ground) state of \iso{15}{C} will be populated~(Fig.~\ref{fig1}). 
If the \iso{16}{N} isomer is also a neutron halo state as in the case of \iso{15}{C}, the ratio in Eq.~\ref{eq:1} should be close to unity, because the overlap of the neutron wavefunction $\beta^m(1s_{1/2})$ should be close to unity.
A shell-model calculation using the WBP interaction~\cite{WBP} which confined the protons to the $p$-shell and the neutrons in the $sd$ shell, predicted that $R$ is indeed 1 and that the occupancies and spatial distributions of the two $\nu$\orb{1}{s}{1} orbitals are almost identical. However, the shell-model calculation may only be partially correct because it may not treat the extended spatial behaviour of a neutron halo wavefunction properly~\cite{Otsuka1993}.

\begin{figure}[h]
\centering
\begin{subfigure}
\centering
\includegraphics[trim=8.2cm 8.9cm 8.2cm 6.0cm, clip,width=8.5cm]{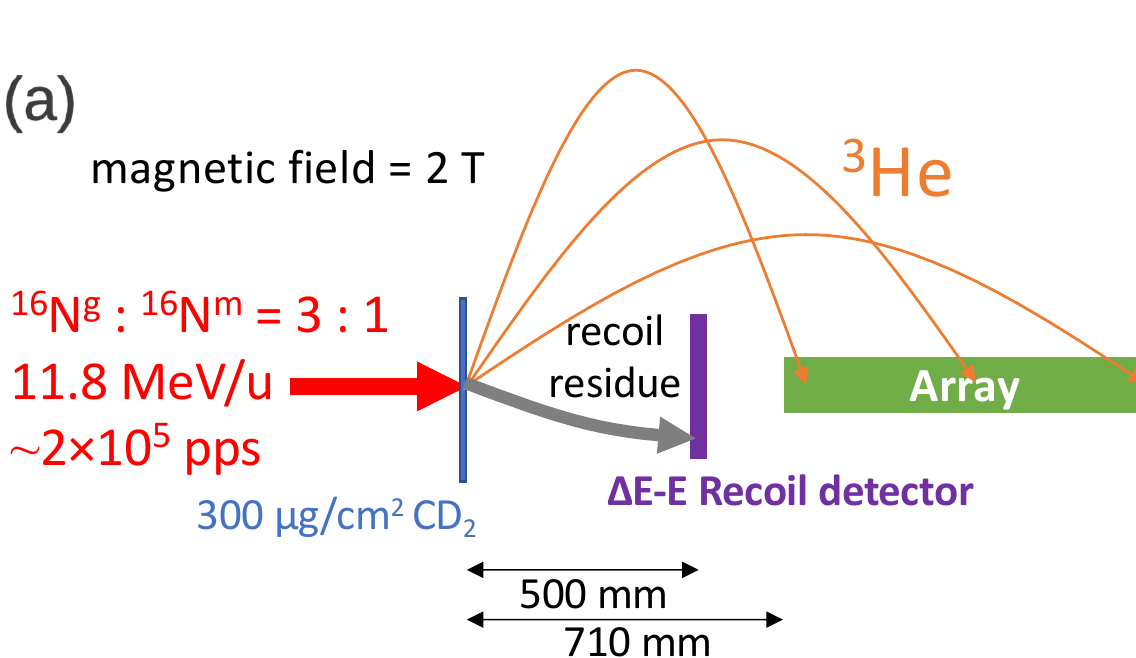}
\end{subfigure}
\begin{subfigure}
\centering
\includegraphics[trim=1cm 1.2cm 2cm 2cm, clip,width=8.5cm]{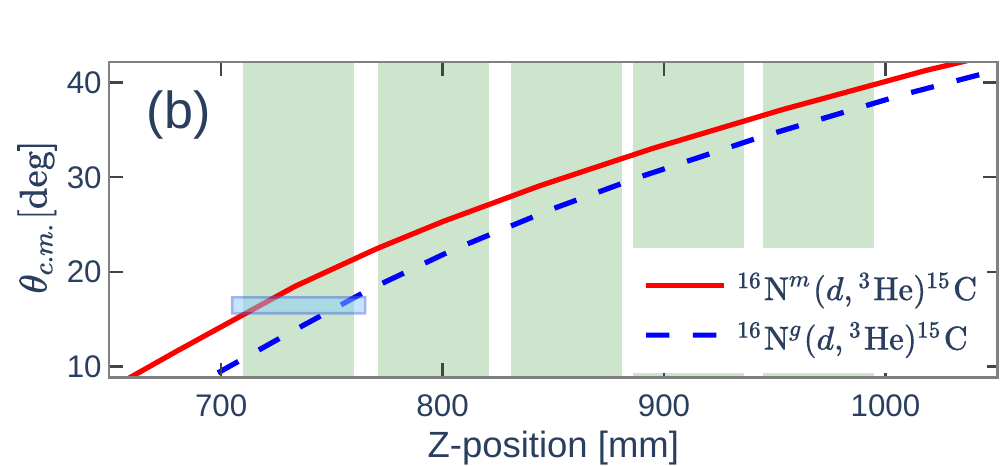}
\end{subfigure}
\caption{\label{fig2} (a) Illustration of the experimental setup. See the main text for additional details. (b) The center-of-mass angle $\theta_{c.m.}$ vs Z-position. The green bands indicate the detector positions of the Si array. The blue square box indicates the common $\theta_{c.m.}$ region for the \dHe~reactions.}
\end{figure}

\section{Experimental setup}

The simultaneous $^{16}$N$^{g,m}$($d$,$^3$He) reactions were performed at Argonne National Laboratory using the ATLAS in-flight system~\cite{RAISOR} and the HELIOS spectrometer~\cite{Lighthall10}. 
The \iso{16}{N} secondary beam, at an energy of  11.8~MeV/$u$ was produced from a \iso{15}{N} primary beam at 13.5~MeV/$u$  bombarding a cryogenically cooled ($\sim$90~K) deuterium gas cell at \mbox{1400~mbar}~\cite{gascell}. 
The total intensity of the \iso{16}{N} secondary beam (ground state and isomeric state) was $\approx 2\times 10^{5}$~pps, with primary beam current of $\approx 150$ particle nano-Ampres on the production target. 
The purity of \iso{16}{N}$^{g,m}$ beam with respect to other isotopes was $> 50\%$, with the main contaminant stemming from the unreacted \iso{15}{N} beam.
Beam identification, rate, and purity were determined using a lower primary beam intensities with a silicon $\Delta E-E$ telescope ($\Delta E = 60~\mu$m and $E=1000~\mu$m) located at the reaction target position inside HELIOS.

A separate experiment was carried out to measure the \iso{16}{N}$^m$ content in the beam~\cite{RAISOR}. 
From that work, the \iso{16}{N}$^m$  component was determined to be 24(2)\% of the total \iso{16}{N} beam at an energy of 11.8~MeV/$u$. 
The isomer fraction was deduced through a comparison between the number of 120-keV $\gamma$-rays decaying from \iso{16}{N}$^m$ and the number of \iso{16}{N}$^g$ ground state decays. 
The uncertainty on the isomer fraction is primarily due to the uncertainties on the $\beta-$delayed $\gamma$-ray branches in the \iso{16}{N} ground state decay (see Ref.~\cite{RAISOR} for additional details). 
The isomer fraction measurement physically took place at the entrance of the HELIOS spectrometer to ensure the same beam-line acceptance as in the present work. In addition, the primary beam, production target, and ATLAS in-flight system properties also mirrored those of the \iso{16}{N}$^{g,m}$($d$,\iso{3}{He}) data collection at 11.8~MeV/$u$. 
The properties of the primary beam and the production target remained stable throughout the present work.
Therefore, the isomer fractions were also assumed to remain constant throughout.



The \iso{16}{N} secondary beam, mixed with both ground state and isomeric state, was transported to HELIOS, a solenoid spectrometer which had a 2~T magnetic field (Fig.\ref{fig2}). The beam bombarded 
a CD$_2$ target of thickness $\approx300~\mu\textrm{g/cm}^2$. The heavy-ion recoil particles, such as $^{15}$C and $^{15}$N, were detected and identified using quadrant silicon $\Delta$E-E recoil detectors %
placed at \mbox{500 mm} downstream of the target position with %
dimensions of 10~mm inner radius, 50~mm outer radius, and placed 10 mm apart. The thickness of the $\Delta E$ and E detectors was 500 $\mu$m and 1000 $\mu$m, respectively. The light 
charged particle \iso{3}{He} was detected using a position-sensitive Si-detector array 
with the first active area beginning at 710 mm downstream of the target position. The array had 6-sides in a hexagonal shape with 5 detectors on each side. Position is determined by resistive charge division along the beam direction : the active area was 50 mm (along the beam axis) $\times$ 10 mm with a position resolution of roughly 1 mm along the beam axis. The distances of the padding and spacing between detectors in the same side of the array were \mbox{$\sim$8.75 mm}. Therefore, the total length covered by the array was 285 mm. The corresponding angular acceptance in the center-of-mass frame was from 11$^\circ$ to 38$^\circ$ for the \iso{16}{N}$^g\rightarrow$\iso{15}{C}(0.75) reaction and from 16$^\circ$ to 40$^\circ$ for the \iso{16}{N}$^m\rightarrow$\iso{15}{C}(0.00) reaction (Fig.~\ref{fig2}).
\begin{figure}[h]
\centering
\includegraphics[trim=0.6cm 1.0cm 0 2.5cm, clip, width=10cm]{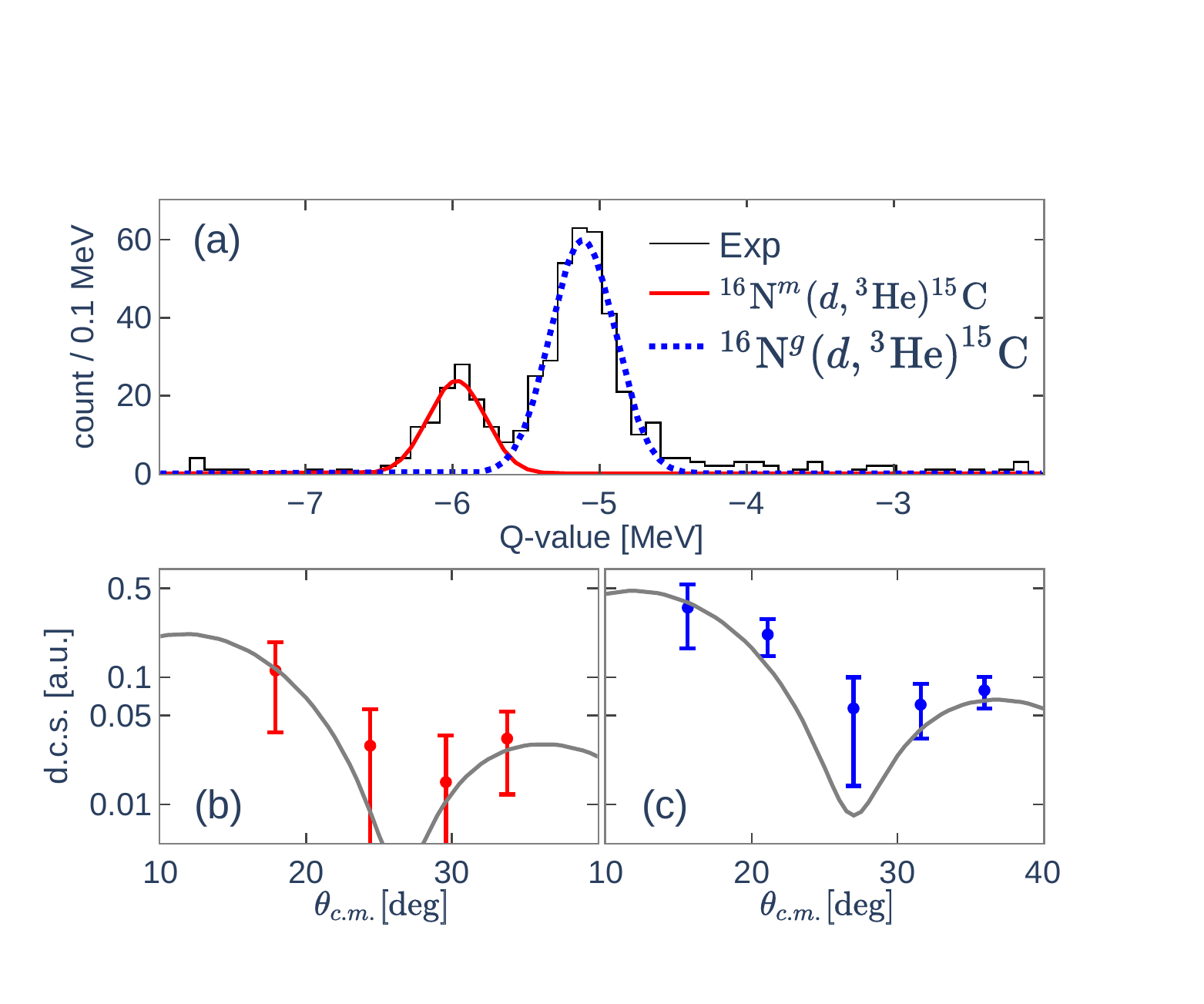}
\caption{\label{fig3} The $^{16m,g}$N($d$,$^3$He) experimental $Q$-value spectrum and angular distributions with gating on the coincident time for \iso{3}{He} and the \iso{15}{C} was identified by the recoil detector. The peaks and differential cross sections have not been normalized by the relative isomer to ground state content of the beam. (a) The $Q$-value of the \iso{16}{N}(iso)$\rightarrow$\iso{15}{C}(0.00) reaction is -5.985 MeV, and that of the \iso{16}{N}(g.s.)$\rightarrow$\iso{15}{C}(0.74) reaction is -5.125 MeV~\cite{Tilley1993, Wang2021}. The peaks are fitted with Gaussian distributions and the FWHM is $\sim500$ keV. The reduced $\chi^2$ of the fitting was 1.54. (b) and (c) The angular distributions for the two final states were fitted with an $\ell=1$ $p$-wave angular distribution calculated from the DWBA approach (see text). 
}
\end{figure}

\section{Experimental Result}
The reaction channels of ($d$,\iso{3}{He}) were identified using the coincidence 
timing between the Si array and the recoil detector telescope as well as the isotope identification of the heavy-ion recoil. 
The resolution of the relative timing between the Si array and the recoil detector telescope was $\approx3$~ns.
The light, charged particles undergo a cyclotron motion within the HELIOS magnetic field with a period depending on the mass-to-charge ratio. 
The coincident time difference %
between the array and the recoil detector 
for \iso{3}{He} is $\sim$37 ns, while that for protons, deuterons, and $^3$H, are $\sim$20 ns, $\sim$55 ns, and $\sim$87 ns, respectively. 
As such, each reaction channel can be uniquely defined from the coincidence time and the recoil gating.



The experimental $Q$-value spectrum for the two ($d$,$^{3}$He) reactions is shown in Fig.~\ref{fig3}a. 
The left smaller peak %
at around $-6$~MeV comes from the \iso{16}{N}$^{m}$($d$,\iso{3}{He})\iso{15}{C}$_{g.s.}$ reaction and the right larger peak 
near $-5.2$~MeV comes from the \iso{16}{N}$^{g}$($d$,\iso{3}{He})\iso{15}{C}(0.74) reaction. The $Q$-values for these two reactions are different by 0.86 MeV, which is the sum of the excited state energy of \iso{15}{C}(0.74) and the  isomeric state energy of \iso{16}{N}$^m$(0.12). The $Q$-value resolution was $\sim 500$~keV FWHM. 
The main contributors to the resolution include energy losses and straggling in the $\sim 300$~$\mu$g/cm$^2$ target and the emittance properties of the in-flight radioactive beam.

The DWBA calculations were carried out using the software code PTOLEMY~\cite{Ptolemy}. The bound-state form factors were taken from Ref.~\cite{Ben2013}. The optical-model potentials for the deuteron incoming channel and the \iso{3}{He} outgoing channel were taken from Ref.~\cite{AnCai} and Ref.~\cite{Xu2011} respectively.  Various combinations of global potentials were calculated (deuteron~\cite{AnCai, Han2006, Daehnick1980}, \iso{3}{He}~\cite{Xu2011, Liang2009, Pang2009}) and used for estimating the uncertainty from the DWBA calculations.

The angular distributions of the two peaks~(Fig.~\ref{fig3}b and 3c) were fitted with an $\ell=1$ \orb{0}{p}{1} angular distribution from the DWBA calculations. Spectroscopic factors were extracted from fits of the angular distribution to the experimental data, after a correction for the beam composition had been applied. Each detector group on the array, spanning 50 mm, was treated as one $\theta_{c.m.}$ angular bin. The SF of the \iso{16}{N}(g.s)$\rightarrow$\iso{15}{C}(0.74) reaction was normalized to 1, resulting in a relative SF of 1.00$\pm$0.17. Using the same normalization, the \iso{16}{N}(iso.)$\rightarrow$\iso{15}{C}(0.00) SF was 1.23$\pm$0.20, and \mbox{$R$ was $0.81\pm0.19$}~(Eq.~\ref{eq:1}). The uncertainty of $R$ mainly came from the statistics ($0.18$) in the fitting of the differential cross section~(Fig.~\ref{fig3}) which is larger than that from the isomer content ($0.06$) and that from the DWBA calculation ($0.04$).


Alternatively, $R$ can be determined by taking the integrated yields over the entire angular coverage~(Fig.~\ref{fig3}a), correcting for the isomer content in the beam, and normalizing to the integrated DWBA cross sections over the same ranges, the ratio $R$ was deduced to be \mbox{$0.92\pm0.15$}. Contributions to the uncertainty from the statistics ($0.14$), the isomer content in the beam ($0.06$), and due to the the DWBA calculations (\mbox{$\approx0.03$}), were included. In addition, the \iso{16}{N}$^g$($d$,\iso{3}{He}) and \iso{16}{N}$^m$($d$,\iso{3}{He}) reactions share a common center-of-mass angle $\theta_{c.m.}$ from  $15.63^\circ$ to $17.30^\circ$ over the same most upstream detectors in the Si array (blue shaded box in the lower part of Fig.~\ref{fig2}). 
The relative SF over this limited angle range was $R = 0.75\pm0.18$ 
again after correcting for the isomer content in the beam, and normalizing to the integrated DWBA cross sections. 
The ratio of the DWBA cross sections over this limited angular range was $0.85\pm0.03$. Therefore, the uncertainty from the DWBA calculations was only 4\%, and much smaller than the statistical uncertainty of 24\%. Although other Si array detector positions also had common $\theta_{c.m.}$ coverage, they did not have sufficient statistics for a similar analysis. The extraction of $R$ from both integration ranges, as well as the fit to the angular distributions, yields an average value of $R=0.83\pm0.17$. 

\section{Discussion}

The ratio $R$ extracted from the $^{16}$N$^{m,g}$($d$,\iso{3}{He}) reaction data was determined to be $0.83\pm0.17$. 
As discussed above, from Eq.~\ref{eq:1}, the ratio should be $\gtrsim 1$ under the assumptions laid out there. 
Namely, that the $\ell=1$ proton single-particle overlaps should be near-identical for the two reactions, and that the $\ell=2$ neutron wavefunctions are also the same between the $^{15}$C($5/2^+$) and $^{16}$N($2^-$) states, i.e. $\beta^g(0d_{5/2}) \approx 1$. 
Thus, consistency of the extracted value of $R$ with Eq.~\ref{eq:1} suggests that there are no large differences between the \orb{1}{s}{1} neutrons in both the \iso{15}{C}($1/2^+$) and \iso{16}{N}$^m$($0^-$) levels, namely $\beta^m(1s_{1/2}) \approx 1$. 
Given that the \iso{15}{C} ground state is a well-established neutron-halo state~\cite{Fang2004}, the present result supports the labelling of the isomeric state in \iso{16}{N} as an excited neutron-halo state. 
This is also consistent with the conclusions drawn in Refs.~\cite{Li2016, Fukutome2022}.

\begin{figure}[h]
\centering
\includegraphics[width=8.5cm]{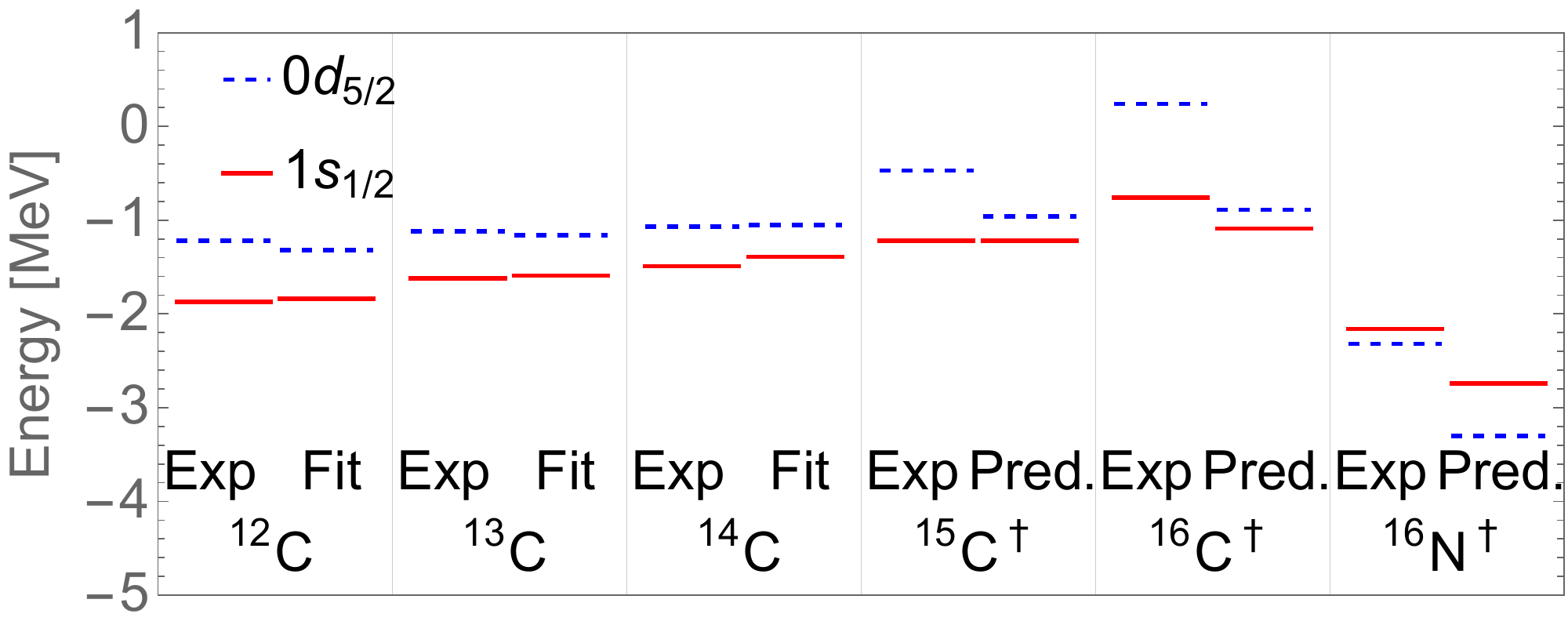}
\caption{\label{fig_ESPS} Effective neutron single-particle energies for the 0$d_{5/2}$ and 1$s_{1/2}$ orbitals taken from Table~\ref{table:EnergyLevel}.}
\end{figure}

A calculation using a Woods-Saxon (WS) potential~\cite{Schwierz2007} was used to better
understand and describe the present result and other observables in various $Z=6$ and $7$ isotopes. %
The WS potential model had fixed parameters of $r_0 = 1.25$~fm, $a_0=0.67$~fm, $r_{SO}=1.20$~fm, and $a_{SO}=0.67$~fm. The spin-orbital strength $V_{SO}$ and central potential depth $V_0 = V + V_{sym}(N-Z)/A$ parameters were extracted from a fit to the empirical ESPEs. The fitting goal was to minimize the quantity $\sum_{i} \left( (e_i - f_i)/e_i \right)^2, $where $e_i$ is the experimental ESPEs and $f_i$ is the WS single-particle energy. %
The normalization to the binding energy was included in order to compensate for the experimental uncertainty of the $p$-shell ESPEs. The empirical neutron ESPEs 
were extracted using the method in Ref.~\cite{Baranger1970}. %
The \orb{0}{p}{3}, \orb{0}{p}{1}, \orb{1}{s}{1}, and \orb{0}{d}{5} orbitals could be determined from the available \iso{12, 14}{C} data \cite{Kaline1971, nucleiA13_15}, while the latter two orbitals only 
were extracted from \iso{13,15,16}{C} and \iso{16}N data~\cite{Bardayan2009, nucleiA13_15, Wuosmaa2010, Guo2014}. Table~\ref{table:EnergyLevel} lists the resulting experimental ESPEs, and Fig.~\ref{fig_ESPS} shows the s-d shell energy levels from Table~\ref{table:EnergyLevel}.

\begin{table}[ht]
\caption{\label{table:EnergyLevel} Effective neutron single-particle energies. The energies in the bracket are from calculations using the WS model. The WS parameters are $V=-57$~MeV, $V_{sym}=45$~MeV, $r_0=1.25$~fm, $a_0=0.67$~fm, $V_{SO}=22$~MeV, $r_{SO}=1.20$~fm and $a_{SO}=0.67$~fm. The energy levels of \iso{15,16}{C} and \iso{16}{N} (marked by $\dagger$) are predicted by the WS model.}
\newcommand\T{\rule{0pt}{3ex}}
\newcommand \B{\rule[-1.8ex]{0pt}{0pt}}
\begin{ruledtabular}
\begin{tabular}{c c c  c  c  c  c }
Orbital & \iso{12}{C}  &  \iso{13}{C} & \iso{14}{C} & \iso{15}{C}$^\dagger$ & \iso{16}{C}$^\dagger$ & \iso{16}{N}$^\dagger$ \\
\hline
\\[-0.7em]
\orb{0}{p}{3} & -16.36  &       & -16.93  &      &  &\\
              & (-15.81)&       & (-14.16)& & & \\
\orb{0}{p}{1} & -9.42   &       & -8.83   &      &  &\\
              & (-12.00)&       & (-10.63)& & & \\ 
\orb{1}{s}{1} & -1.87   & -1.62 & -1.49   & -1.22  & -0.76  & -2.16 \\
              & (-1.84) &(-1.59)&(-1.39)  &(-1.22) &(-1.09) &(-2.74) \\ 
\orb{0}{d}{5} & -1.22   & -1.12 & -1.07   & -0.47  & +0.24  & -2.32 \\
              & (-1.32) &(-1.16)&(-1.05)  &(-0.96) &(-0.89) &(-3.30)\\
 \end{tabular}
 \end{ruledtabular}
 \end{table}
 
Only the \iso{12,13,14}{C} energy levels were included in the fit. The best-fit parameter set was $V=-57$~MeV, $V_{sym}=45$~MeV, and $V_{SO}=22$~MeV. The calculated single-particle energies resulting from the best-fit parameter set are listed in parentheses in Table~\ref{table:EnergyLevel}. 
One major deviation in comparison with data comes from the absolute energies of the deeply bound $p$-shell energies. However, as with the deviations among the s-d shell results, the rms deviations are 63~keV. The single-particle energies of \iso{15}{C}, \iso{16}{C} ~\cite{Tilley1993} and \iso{16}{N}~\cite{Wuosmaa2010, Tilley1993} were calculated from the best-fit parameters and the results are in good agreement with the experimental ESPEs.

Although the WS model is not meant for precise global predictions, the reasonable reproduction of  the ESPEs in the region for \iso{15,16}{C} and \iso{16}{N} shows that the model is more than valid and appears to capture most of the key components of the interactions determining the binding energies. For instance, the \orb{0}{d}{5} orbital is more bound than the \orb{0}{s}{1} orbital in \iso{16}{N} from our WS calculation, in agreement with the experimental data~(Table~\ref{table:EnergyLevel}). This suggests the change of orbital ordering requires no additional interactions to reproduce this behavior. In fact, the change of orbital ordering between the $s$- and $d$-orbitals is also a consequence of the weak-binding effect~\cite{Hoffman2014, Hoffman2014_2, Hoffman2016}. Of course, a microscopic description of the change of the shell ordering has been around for some time as described in Ref.~\cite{Talmi1960}.

\subsection{Root-mean-square and matter radius}

From the Woods-Saxon calculations, the rms radii of the s-orbitals corresponding to the \iso{15}{C} ground state and \iso{16}{N} isomeric state are 5.5 fm and 4.5 fm, respectively.
Since the value of $R$ should be $\gtrsim 1$, and the upper limit of the measured value of $R$ is 1.00, by setting $\beta^m(1s_{1/2})$ to be 1.00, and 
adjusting the WS potential depth $V$ (from -46 MeV to -58 MeV) of \iso{16}{N} to vary the rms radius of the s-orbital, the overlap integral is 1.0 when the rms radius is 4.8 fm. This agrees with the rms radius extracted using ANC~\cite{Li2016}. At this limit, the $V$ is \mbox{-52~MeV} and the ESPE of the \orb{1}{s}{1} is \mbox{-2.16~MeV}. The best fit parameters (see Table.~\ref{table:EnergyLevel}) are used in the rest of the discussion.
 
\begin{table*}[ht]
\caption{\label{table:MatterRadii} Matter radii of some carbon and nitrogen isotopes. The matter radius of \iso{14}{C} is assumed to be 2.30 fm. The matter radii of \iso{12,13}C were calculated by the same recipe in Ref.~\cite{Fortune2016} but using hole-state. The theoretical results are marker by $\dagger$.
}
\newcommand\T{\rule{0pt}{3ex}}
\newcommand \B{\rule[-1.8ex]{0pt}{0pt}}
\begin{ruledtabular}
\begin{tabular}{c  c  c  c  c  c  }
Nucleus & Configuration & Ref.\cite{Ozawa2001}  & Ref.\cite{Ritu2016} & Ref.\cite{Fortune2016}$^\dagger$ & WS$^\dagger$  \\
\hline
\\[-0.7em]
\iso{12}{C} &        $\nu(1p_{1/2})^{-2}$             & 2.33(2)  & 2.35(2)  & & 2.14  \\
\iso{13}{C} &       $\nu(1p_{1/2})^{-1}$            & 2.28(4)  & 2.28(4) & & 2.24\\
\iso{14}{C} & core                     & 2.30(7) & 2.33(7) & (2.30) & (2.30) \\
\iso{15}{C} & $\nu(1s_{1/2})$                      & 2.40(5) & 2.54(4) & 2.62   & 2.62 \\
\iso{16}{C} & $\nu(1s_{1/2})^2$  &  &  &  & 2.90 \\
\iso{16}{C} & $\nu(0d_{5/2})^2$  &  &  &  & 2.54 \\
\iso{16}{C} & 54\%$\nu(1s_{1/2})^2$ + 46\%$\nu(0d_{5/2})^2$  & 2.70(3) & 2.74(3) & 2.62   & 2.70 \\
\iso{16}{N}$^g$ & $\pi(1p_{1/2})\nu(0d_{5/2})$   &2.50(10) &  &  & 2.40 \\
\iso{16}{N}$^m$ & $\pi(1p_{1/2})\nu(1s_{1/2})$  &  &  &  & 2.49 \\
 \end{tabular}
 \end{ruledtabular}
 \end{table*}
 
\begin{figure}[h]
\centering
\includegraphics[trim=0.7cm 0.9cm 0 2.5cm, clip, width=10cm]{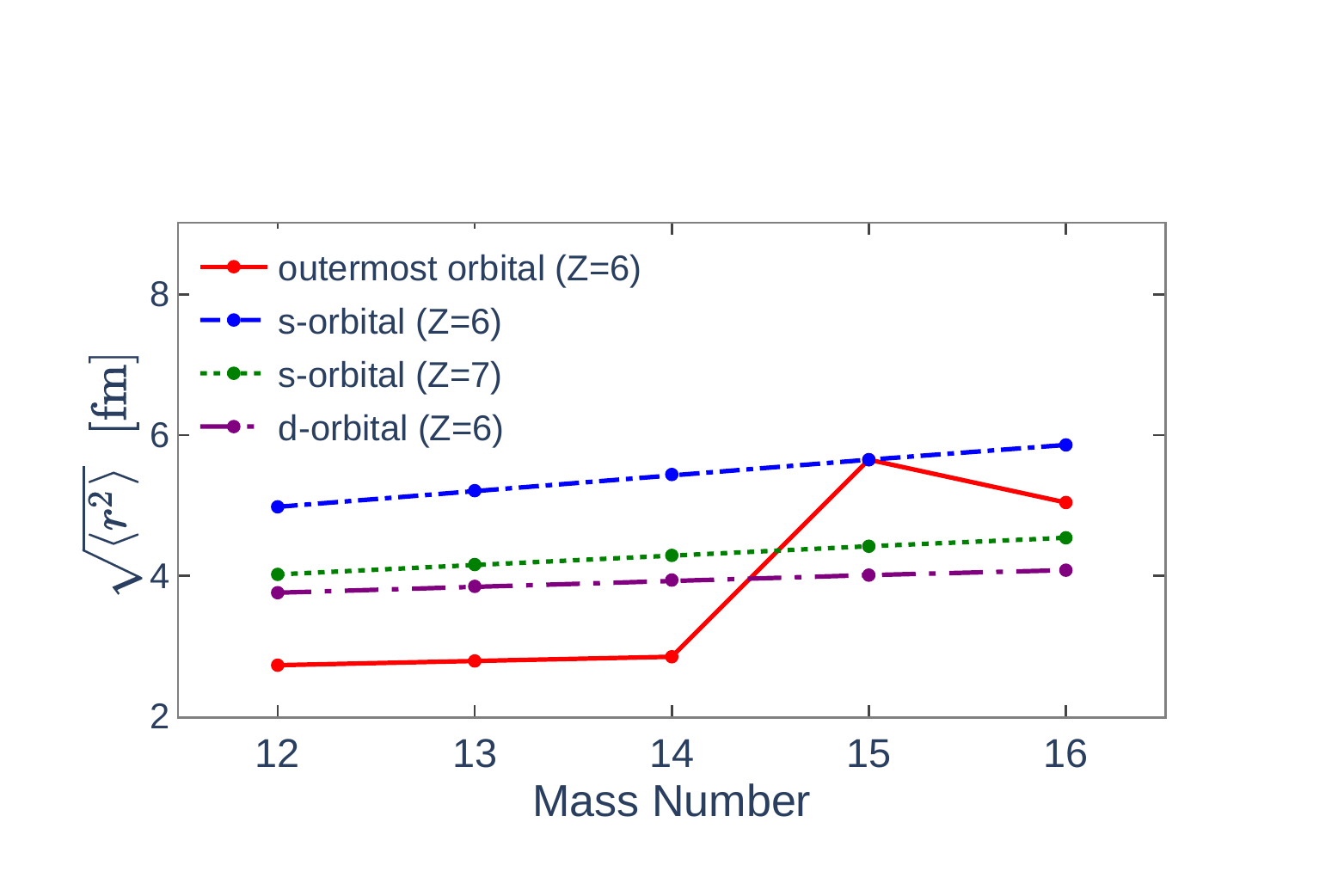}
\caption{\label{fig_rms} The orbital rms radii of carbon and nitrogen isotopes from the WS calculation with the parameter set given in the main text. The outermost neutron orbital for \iso{12-14}{C} is $p$-orbital. The configuration of 54\% $s^2$ + 46\% $d^2$ is being used for the outermost orbital in \iso{16}{C}~\cite{Fortune2016}.}
\end{figure}

Fig.~\ref{fig_rms} shows the orbital rms radii based on the WS calculation for the outermost (\orb{1}{s}{1} and \orb{0}{d}{5}) neutron orbitals of the carbon isotopes as well as the \orb{1}{s}{1} neutron orbital for the nitrogen isotopes. 
The $54\%s^2+46\%d^2$  configuration of the \iso{16}{C} ground state is being used \cite{Fortune2016}, and a pure configuration is taken for the others. 
There is an expected similarity between the trend in the calculated orbital rms radii and the trends of the extracted and calculated matter radii of Table~\ref{table:MatterRadii}, emphasizing the role that the outermost orbital plays in defining the spatial extent of these systems. 
Due to the direct relation between a reaction cross section and the spatial extent of the nucleus, it is also not surprising that the reaction cross section data measured up to $\sim$100$A$~MeV ~\cite{Fang2004,Villari1991,Ozawa2001} also follows a similar trend. 
It should be noted that the orbital rms radii of the s-orbital in \iso{12-16}{C} are all $\sim$5.5 fm as they have similar (weak) binding energies (Table~\ref{table:EnergyLevel}).
This general behavior is a consequence of the weak-binding effect~\cite{Hoffman2014, Hoffman2014_2, Hoffman2016}. 
Particularly, the excited $1/2^+$ state in \iso{13}{C} should also display neutron-halo features as it has a relatively pure \iso{12}{C}$\otimes(\nu1s_{1/2})$ configuration and a neutron $s$-orbital rms radius of $\sim5.2$~fm~\cite{Belyaeva2014, Liu2001}. 
If the excited state of \iso{13}{C} is not relatively pure, like the state of \iso{16}{C}, the excited state of \iso{13}{C} would not be an excited halo.

Table~\ref{table:MatterRadii} lists the matter radii that were calculated using the recipe from Ref.~\cite{Fortune2016}. Our results from the WS model calculations show essentially no significant difference with Ref.~\cite{Fortune2016}, which is not surprising considering the similarity between the two WS models. If the configuration for \iso{16}{C} is \mbox{30\%$s^2$+70\%$d^2$}, as determined in Ref.~\cite{Wuosmaa2010}, the matter radius of \iso{16}{C} is 2.64 fm, closer to the value given from Ref.~\cite{Fortune2016}, but away from the experimental values. The matter radius of \iso{15}{C} is still different from the experimental value \cite{Ritu2016}. The rms radius of the \orb{1}{s}{1} neutron deduced using the experimental values from Ref.~\cite{Ritu2016} is 4.72$\pm$0.16 fm, which is $\sim$1~fm smaller than that from our WS calculation, but still much larger than the size of \iso{15}{C}, which is $\sim1.25A^{1/3}=3.08$~fm.





\begin{figure}[h]
\centering
\includegraphics[trim=0.cm 0.9cm 0.0cm 2.6cm, clip, width=9.8cm]{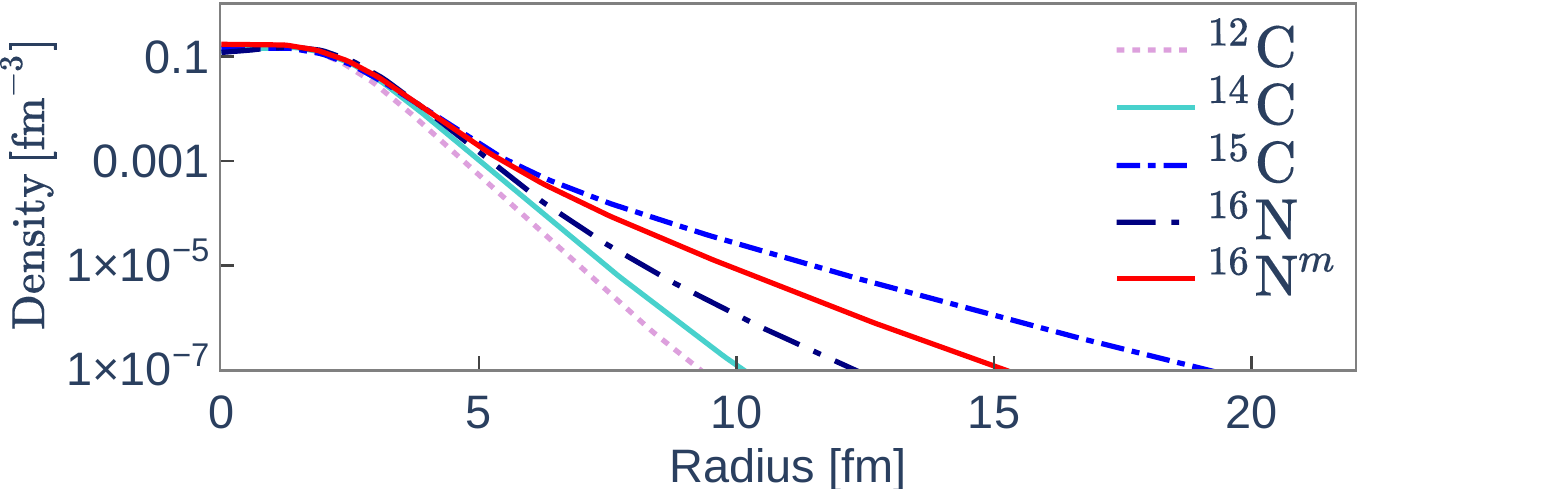}
\caption{\label{fig_WSdensity} The density distributions of \iso{12}{C}, \iso{14}{C}, \iso{15}{C}, \iso{16}{N}, and \iso{16}{N}$^m$ from the Woods-Saxon calculation [Table.~\ref{table:EnergyLevel}]. The tails of the density functions of \iso{12,14}{C} are from the $p$-orbital, and that of \iso{15}{C} and \iso{16}{N}(iso) are from the $s$-orbital, and that of \iso{16}{N} is from the $d$-orbital. The energy level of the \orb{1}{s}{1} \iso{16}{N}$^{m}$ is lower then that of \iso{15}{C}, so that the density is not as extended as \iso{15}{C}. The $d$-state in \iso{16}{N} is much less bound than the $p$-state in \iso{12,14}{C}, by $\sim 12$ MeV (Table~\ref{table:EnergyLevel}), which makes it slightly more extended.} 
\end{figure}



\subsection{Density function and reaction cross-section}

The wavefunctions from the WS model were also used to calculate the density function of each state by summing all occupied orbitals and then normalizing to the mass number. Fig.~\ref{fig_WSdensity} shows the resulting density functions for the states of interest. The \iso{15}{C} ground state and \iso{16}{N}$^{m}$ extend with the longest tails as compared with \iso{12,14}{C}, as expected. The tails of \iso{14,15}{C} are similar to those shown in Ref.~\cite{Fang2004}. 
Notice that the binding energy of \iso{16}{N} is -2.74 MeV in our WS model, which is 0.58 MeV more bound than the experimental value. The density function for this level continues to extend further when empirical binding energy is used. 

\begin{figure}[h]
\centering
\includegraphics[trim=0.0cm 1.0cm 0cm 2.5cm, clip, width=10cm]{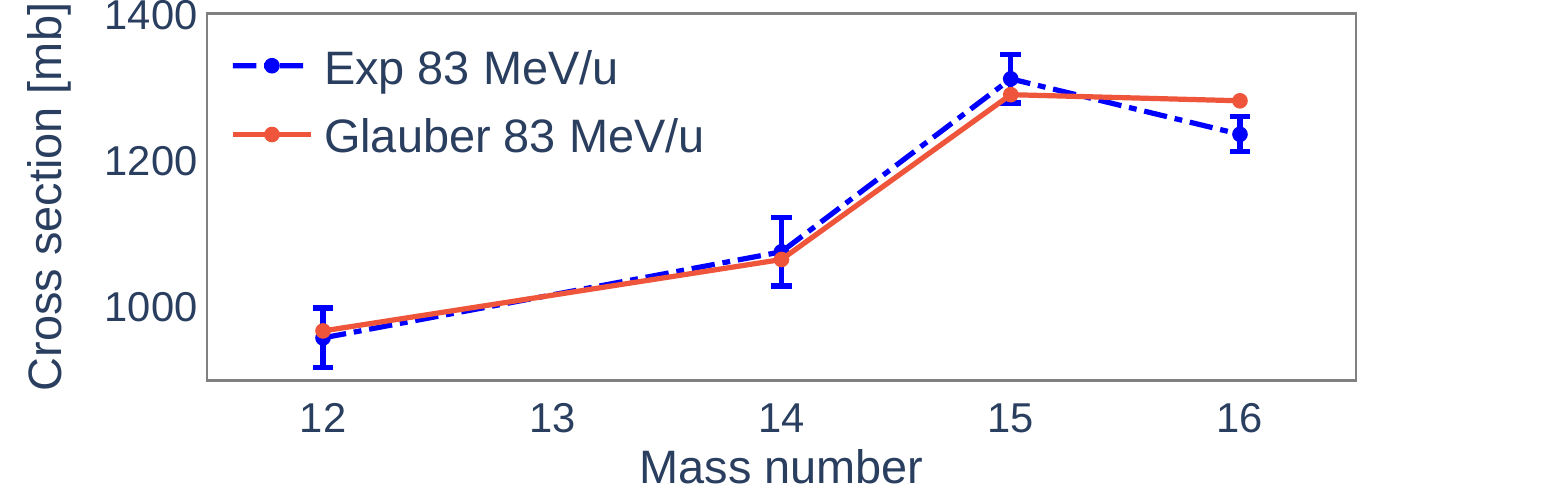}
\caption{\label{fig_GM} The reaction cross section of \iso{A}{C} + \iso{12}{C} at 83 MeV/$u$. A zero-range Coulomb-corrected Glauber model~\cite{Charagi1990} was used to calculate cross section from the density distributions from Fig.~\ref{fig_WSdensity}. The experimental cross sections were extracted from Ref.~\cite{Fang2004}. The configuration of $54\%s^2 + 46\%d^2$ is used in the case of \iso{16}{C}.}
\end{figure}

The density functions were also used in a zero-range Coulomb-corrected Glauber-type calculation~\cite{Charagi1990} to calculate reaction cross sections. The nucleon-nucleon reaction cross section was taken from Ref.~\cite{Charagi1990}. 
The results reproduce the experimental values as shown in Fig.~\ref{fig_GM}. In particular, the calculation shows that the tail of the density distribution of \iso{15}{C} is responsible for the enhancement of the reaction cross section. 

\subsection{Momentum distribution}

As another demonstration of the validity of the WS model to the states of interest, the longitudinal momentum distribution of the \orb{1}{s}{1} orbital of \iso{15}{C} was calculated. The FWHM width of the \orb{1}{s}{1} orbital was calculated to be 65 MeV/c from our WS model, which is consistent with the experimental width of $71\pm9$ MeV/c for the \iso{15}{C}$\rightarrow$\iso{14}{C} reaction in Ref.~\cite{Fang2004}, that was carried out at 83 MeV/u. 
The calculated FWHM width of the \orb{0}{d}{5} orbital is 270 MeV/c.
The experimental resolution could make the experimental distribution larger. Ref.~\cite{Fang2004} used a hybrid density function that combines the harmonic oscillator and a long Yukawa tail to reproduce the experimental result. It patches up the harmonic oscillator and echoes the long tail from the WS model when the binding energy is weak. The long Yukawa tail is also in line with the modified wavefunctions in Ref.~\cite{Otsuka1993}.

\subsection{A general consideration on neutron halo states}

In each of the carbon isotopes discussed here, the $1s$-orbitals is weakly bound (Table.~\ref{table:EnergyLevel}). Similarly, the $1s$-orbitals in $Z=7$ nuclei move into this energy region for $A \sim 15-16$. All weakly-bound $1s$-orbitals will have large rms radii, independent of whether they are the ground or an excited state. 
The rms radius of \iso{16}{C} is then slightly reduced due to the mixing between the $s$- and $d$-orbitals~\cite{Wuosmaa2010}. Considering that the $s$-orbital is only slightly more bound in \iso{16}{N} than that in \iso{15}{C}, it also has a relatively large rms radius of \mbox{4.60~fm} from the WS calculation that agreed with the rms radius extracted using the ANC~\cite{Li2016}. Furthermore, given the similarities of neutron \orb{0}{s}{1} wavefunctions of \iso{16}{N}$^m$ and \iso{15}{C}, it is expected that the radial overlap between the two should be close to 1, consistent with the $R$ value extracted in the present work.

The $1^-$ excited state of \iso{16}{N} will also have the same structure as the isomer state as they both originate from the coupling of the $\pi$\orb{0}{p}{1} and $\nu$\orb{1}{s}{1} orbitals. Hence, the $1^-$ state will be dominated by a neutron $s$-orbital and will posses a large orbital rms radius. Recently, Ref.~\cite{Ziliani2021} measured the excitation energies of the $s$-orbital doublet ($J^\pi=0^-,1^-$) in \iso{18}{N} ($N=11$), which correspond to binding energies of $\approx1.7$ MeV. 
This result was predicted in Ref.~\cite{Calem2013} using the linear two-body matrix argument from Ref.~\cite{Talmi1960}. According to the WS model, and the dominant $s$-component extracted for the $1^-$ level~\cite{Calem2013}, these two excited states in \iso{18}{N} should also be considered as excited neutron halo states.

The prevalence of levels having single neutron-halo properties is emphasized by the commonality between the weakly-bound $s$-states contained within the ground or excited levels of the $Z=6-7$ isotopes. This is further supported by the self-contained description of these halo-like $s$-states including the \iso{15}{C} ground state, as well as other $\ell>0$ states by the WS calculations.
Hence, it can be concluded that the \iso{15}{C} neutron-halo ground state is in fact not a unique level at all, in the sense that no special treatment was required by the WS model 
to fully reproduce its associated observables.

\section{Summary}

The \iso{16}{N}$^{g,m}$($d$,\iso{3}{He}) reactions at 11.8 MeV/$u$ were measured simultaneously, and the ratio of the SFs 
was determined to be $R = 0.83\pm0.17$, and consistent with unity. The simultaneous measurement reduced the systematic uncertainty from the experimental setup and in the DWBA analysis. The main source of uncertainty was from the statistics and the \iso{16}{N} beam composition. The ($d$,\iso{3}{He}) result indicates that the neutron \orb{1}{s}{1} orbitals in \iso{16}{N} and \iso{15}{C} are similar and that the isomeric state of \iso{16}{N} should be considered to be an excited neutron halo state. 
Calculations were carried out using a WS potential model and they were able to reproduce the binding energies and the order of the single-particle orbitals, the radial spreads, the longitudinal momentum distributions, and the reaction cross sections using a zero-range Coulomb corrected Glauber model, for isotopes near the \iso{15}{C} region. Since this region is not too far away from the valley of stability, there is no surprise that the WS model should be able to capture the gross structure \cite{Sherr1996}. Finally, it has been determined that a description of the observables associated with the \iso{15}{C} ground state by the WS model required no special treatment.


\section{Acknowledgement}
The authors would like to acknowledge the hard work of the support and operations staff at ATLAS for providing the \iso{16}{N} isomer beam. This research used resources of Argonne National Laboratory’s ATLAS facility, which is a Department of Energy Office of Science User Facility. This material is based upon work supported by the U.S. Department of Energy, Office of Science, Office of Nuclear Physics, under Contract No. DE-AC02-06CH11357 (ANL), DE-AC05-00OR22725 (ORNL), DE-SC0020451 (NSCL), and Grant No. DE-FG02-96ER40978 (LSU). This work was also supported by the UK Science and Technology Facilities Council Grants No. ST/P004423/1, No. ST/T004797/1
(Manchester) and by the National Science Foundation, No. PHY-2012522 (FSU) and the Hirose International Scholarship Foundation from Japan.


\end{document}